\documentclass[]{tPHM2e}

\begin{document}
\doi{10.1080/14786435.20xx.xxxxxx}
\issn{1478-6443}
\issnp{1478-6435}
\jvol{00} \jnum{00} \jyear{2012} %\jmonth{21 December}

\markboth{Xiao Lin, Sergey L. Bud'ko and Paul C. Canfield}{Philosophical Magazine}

\articletype{}
\title{ Physical properties of single crystalline BaSn$_5$}

\author{Xiao Lin, Sergey L. Bud'ko and Paul C. Canfield\thanks{\vspace{6pt}}\\\vspace{6pt}  
{\em{Department of Physics and Astronomy and Ames Laboratory, Iowa State University, Ames, Iowa 50011, U.S.A.}}\\\vspace{6pt}\received{Month 2012} }

\maketitle

\begin{abstract}
We present a comprehensive study of the binary intermetallic superconductor, BaSn$_5$. High-quality single crystalline BaSn$_5$ was grown out of Sn flux. Detailed thermodynamic and transport measurements were performed to study BaSn$_5$'s normal and superconducting state properties. This material appears to be a strongly coupled, multiband superconductor. $H_{\rm c2}(T)$ is almost isotropic. De Haas-van Alphen oscillations were observed and two effective masses were estimated from the FFT spectra. Hydrostatic pressure causes a decrease in the superconducting transition temperature at the rate of $\approx -0.053$ $\pm$ 0.001 K/kbar.\bigskip

\bigskip

\end{abstract}

\section{Introduction}

To search for new superconductors, one of many ways is to look for compounds that share similar features with the already reported superconductors. On the one hand, BaSn$_5$ has a similar band dispersions near the Fermi level ($E_{\rm F}$) as A15 type superconductors, such as V$_3$Si and Nb$_3$Sn \cite{Fassler2001}. On the other hand, BaSn$_5$ forms in \textit{P6/mmm} structure, a variant of AlB$_2$ structure, the prototype of MgB$_2$ which superconducts at $\sim$ 40 K \cite{Nagamatsu2001, Kwok2003, Canfield2005, Wilke2010}.

The first study of BaSn$_5$ can be traced back to 1979 \cite{Moos1979}, however, only recently has its structure been solved \cite{Fassler2001}. As one of the alkaline earth stannides group of superconductors (for SrSn$_4$, SrSn$_3$, BaSn$_3$ the superconducting transition temperatures are $\sim$ 4.8 K \cite{Hoffmann2003, Lin2011}, $\sim$ 5.4 K \cite{Fassler2000} and $\sim$ 2.4 K \cite{Fassler1997} respectively), BaSn$_5$'s superconducting transition temperature is reported to be $\sim$ 4.4 K \cite{Fassler2001}. So far, only its low temperature and low field magnetization has been characterized on polycrystalline samples \cite{Fassler2001}.

In this article we report the growth of single crystalline BaSn$_5$, and the measurement of its thermodynamic and transport properties. Both the superconducting and normal states are characterized. We also present the effect of pressure on the superconducting properties of BaSn$_5$, and the observation of low temperature de Haas-van Alphen oscillations.

\section{Experimental Details}

Single crystals of BaSn$_5$ were grown out of excess Sn by the high-temperature solution technique \cite{Canfield1992}. Elemental Ba and Sn with an atomic ratio of Ba$_8$Sn$_{92}$ were placed in a 2 ml alumina crucible. A second catch crucible stuffed with silica wool was placed on the top of the growth crucible. Both crucibles were sealed in a silica ampoule under approximately 1/3 atmosphere of high purity argon gas. To prevent oxidization of the growth materials, the packing and assembly of the ampoule was performed in a glovebox with a nitrogen atmosphere. This ampoule was heated up to 700 $^\circ$C, then cooled to 425 $^\circ$C, followed by a slow cool over a period of 40 hours to 270 $^\circ$C, at which temperature the excess flux was decanted from the crystals. Crystals of BaSn$_5$ grown in this manner form in rod-like shape of a few mm in length and sub-mm in the other two dimensions. Due to the samples' air-sensitivity, crystals were kept in the glovebox, and efforts were made to minimize their exposure during measurement.

Powder x-ray diffraction data on both non-oxidized and oxidized sample were collected by a Rigaku Miniflex diffractometer with Cu K$_\alpha$ radiation at room temperature. The diffraction pattern of the non-oxidized BaSn$_5$ was taken from the powder of BaSn$_5$ single crystals which was ground in the glovebox. The sample powder was sealed by Kapton film during the measurement to protect it from oxidization. To study the oxidation effect, a second x-ray diffraction was performed on the same powder after removal of Kapton film and a seven-hour exposure to the air. The lattice constants of non-oxidized BaSn$_5$ were statistically determined by measurements of multiple samples with Si (a = 5.4301 $\AA$) as an internal standard.

Temperature- and magnetic-field dependent dc magnetization data were measured in a Quantum Design MPMS-5 SQUID magnetometer. The ac resistance was measured via a standard four-probe method in a Quantum Design PPMS instrument with the ACT option. Platinum wires were attached to the sample using Dupont 4929 silver paint with the current approximately flowing along the longest dimension (crystal's c-axis). Resistance as a function of temperature was measured at different magnetic fields with field's direction parallel to c-axis and ab-plane respectively. A relaxation technique was applied in the heat capacity measurements in a PPMS instrument. For the measurement of low field dc magnetization under pressure, a commercial, HMD, Be-Cu piston-cylinder pressure cell \cite{web} was used. The highest pressure reached $\sim$ 10 kbar with Daphne oil 7373 as a pressure medium and superconducting Pb as a low temperature pressure gauge \cite{Eiling1981}.

\section{Results and discussion}
\begin{figure}
\begin{center}
\resizebox*{10cm}{!}{\includegraphics{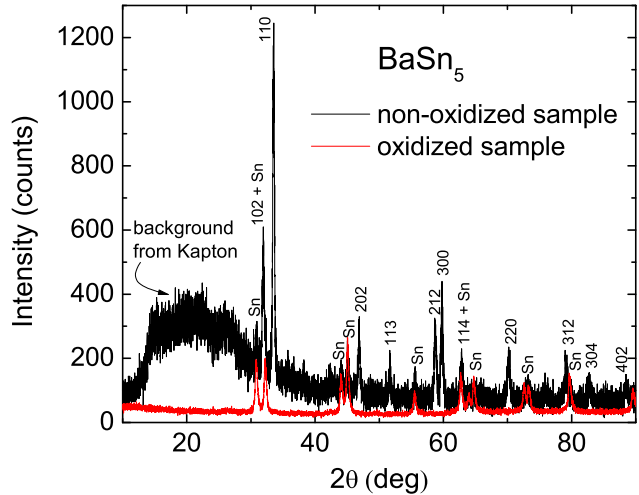}}%
\caption{Comparison of the x-ray patterns taken on non-oxidized and oxidized powdered BaSn$_5$ single crystals. Peaks that belong to BaSn$_5$ are labeled with their \textit{h k l} values. Notes: the only differences between the two runs were (i) removal of Kapton film and (ii) 7 hours exposure to air.}%
\label{X-ray}
\end{center}
\end{figure}

Figure \ref{X-ray} presents the comparison of powder x-ray diffraction on both non-oxidized and oxidized sample. The diffraction pattern from the non-oxidized sample confirms that the synthesized crystals are BaSn$_5$ with \textit{P6/mmm} structure. The obtained lattice parameters are a = 5.368(4) $\AA$, c = 7.097(4) $\AA$, consistent with the reported data \cite{Fassler2001}. Together with BaSn$_5$'s diffraction peaks, several peaks from Sn flux residue are also visible in the diffraction pattern. In contrast, after a seven-hour exposure to air, the same specimen lost all its diffraction peaks of BaSn$_5$. As shown in Fig \ref{X-ray}, only Sn's diffraction peaks survived, with their intensities essentially unchanged. The disappearance of BaSn$_5$ under the powder x-ray diffraction is probably due to the oxidization of BaSn$_5$, resulting in phases that are too small or too disordered to diffract. Similar phenomena was also observed in the powder x-ray diffraction data of non-oxidized and oxidized single crystalline SrSn$_4$ \cite{Lin2011}.

\begin{figure}
\begin{center}
\resizebox*{10cm}{!}{\includegraphics{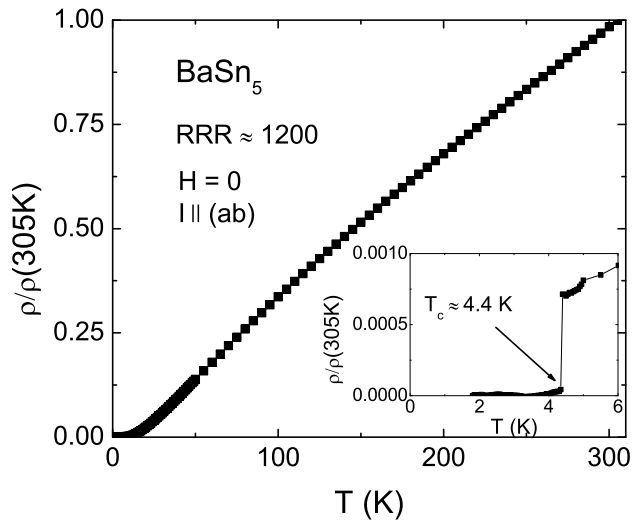}}%
\caption{The temperature-dependent, normalized resistivity of BaSn$_5$. Inset: low temperature data showing the superconducting transition.}%
\label{RT}
\end{center}
\end{figure}

Zero-field, in-plane resistivity of BaSn$_5$ as a function of temperature is presented in Fig \ref{RT}. Due to the sample's irregular shape in cross section and its air-sensitivity, its resistivity is normalized with respect to the room temperature value. To within factor of 25$\%$, the room temperature resistivity reaches approximately 100 $\mu\Omega$ cm. In the higher temperature region, the resistivity manifests a typical metallic behaviour, increasing linearly as the temperature rises. The very substantial residual resistivity ratio (RRR) = $\rho$(305 K) / $\rho$(5.0 K) $\sim$ 1200 indicates that the crystals grow with a very low number of impurities/defects (a conclusion further supported by the observation of quantum oscillations, discussed below). The inset to Fig \ref{RT} shows the low temperature resistivity and a sharp transition to the superconducting state with offset at about 4.4 K, which is consistent with the literature data \cite{Fassler2001}. 

\begin{figure}
\begin{center}
\resizebox*{10cm}{!}{\includegraphics{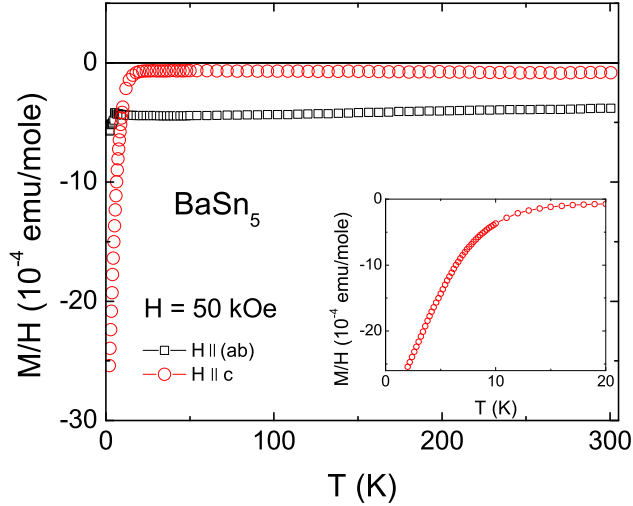}}%
\caption{Anisotropic temperature-dependent magnetic susceptibility, \textit{M/H}, of BaSn$_5$. Inset: enlarged low-temperature part of magnetic susceptibility with \textit{H}$\parallel$c.}%
\label{MT}
\end{center}
\end{figure}

\begin{figure}
\begin{center}
\resizebox*{10cm}{!}{\includegraphics{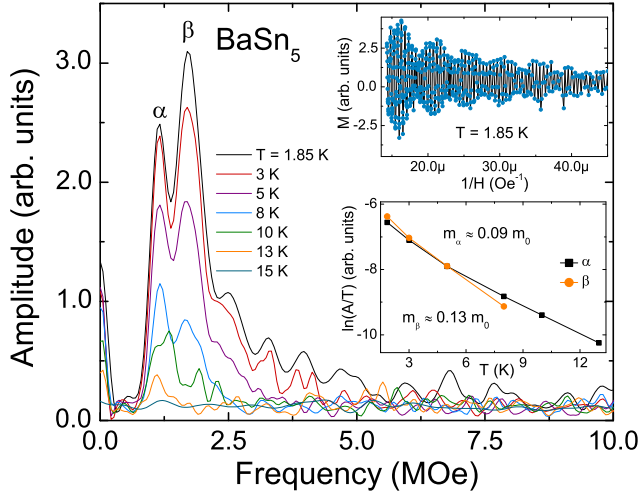}}%
\caption{Fourier spectra of the oscillations in magnetization of BaSn$_5$ up to 15 K. Upper inset: magnetization as a function of inverse magnetic field of BaSn$_5$ at 1.85 K. Lower inset:Temperature dependence of the amplitudes of the observed oscillations.}%
\label{Oscillations}
\end{center}
\end{figure}

The temperature dependent dc magnetic susceptibility, \textit{M/H}, with the magnetic field parallel to c-axis and ab-plane is shown in Fig \ref{MT}. For an applied field of 50 kOe, the normal state of BaSn$_5$ exhibits diamagnetic behaviour in both directions, and essentially does not change with temperature in the higher temperature region. Small anisotropy can be detected above 20 K, with absolute value of $\vert(M/H)_{ab}\vert > \vert(M/H)_c\vert$. However, in the low temperature region, a dramatic enhancement of the diamagnetic feature, especially with field parallel to c-axis, is clearly seen in the inset to Fig \ref{MT}. These sudden changes in \textit{M/H} are most likely brought by de Haas-van Alphen oscillations are shown in the upper inset to Fig \ref{Oscillations}.

To study de Haas-van Alphen oscillations in BaSn$_5$, dc magnetization as a function of applied magnetic field at several different temperatures was measured (Fig \ref{Oscillations}). However, due to the sample's air-sensitivity and irregular shape in ab-plane, only studies with field parallel to c-axis are included in this work. The oscillations in the magnetization can be observed at multiple temperatures, superimposed on the nearly constant magnetic background. The upper inset of Fig \ref{Oscillations} gives an example of these oscillatory behaviours as a function of inverse field up to 70 kOe at 1.85 K. Fast Fourier transform (FFT) was used to convert the oscillations to their Fourier spectra in Fig \ref{Oscillations}. Due to the limitation of signal to noise ratio, only two peaks can be resolved from the spectra with frequencies of 1.16 MG ($\alpha$) and 1.59 MG ($\beta$). Figure \ref{Oscillations} also represents the evolution of the spectra with respect to temperature. It can be clearly seen that the amplitudes of spectra gradually attenuate and finally fade away at about 15 K. For a certain frequency F, the amplitude of the oscillation in the magnetization M is given by the Lifshitz-Kosevitch (LK) equation \cite{Shenberg}:

\[
  M = -2.602\times10^{-6} (\dfrac{2\pi}{HA^{''}})^{1/2} \times \dfrac{GFT\exp (-\alpha px/H)}{p^{3/2}\sinh (-\alpha pT/H)} \times \sin [(\dfrac{2 \pi pF}{H})-1/2 \pm \dfrac{\pi}{4}]
\] where $\alpha$ = 1.47($m/m_0$)$\times$ 10$^5$ G/K, $A^{''}$ is the second derivative of the cross sectional area of the Fermi surface with respect to wave vector along the direction of the applied field, \textit{G} is the reduction factor arising from electron spin, $\rho$ is the number of harmonic of the oscillation, and \textit{x} is the Dingle temperature. Thus, the temperature dependence of the amplitude (\textit{A}) of frequency $\alpha$, $\beta$, plotted in the lower inset to Fig \ref{Oscillations}, can be used to determine the effective mass of the orbits via the LK formula, described above. From the slope of $\ln (A/T)$ plotted as a function of temperature, the effective masses were found to be $m_{\alpha}$ $\approx$ 0.09 $m_0$ and $m_{\beta}$ $\approx$ 0.13 $m_0$, where $m_0$ is the bare electron mass. However, for further understanding of the oscillations and topology of the Fermi surface, angular dependence of the spectra as well as detailed calculations of band structure and Fermi surfaces of BaSn$_5$ are needed.

\begin{figure}
\begin{center}
\resizebox*{14cm}{!}{\includegraphics{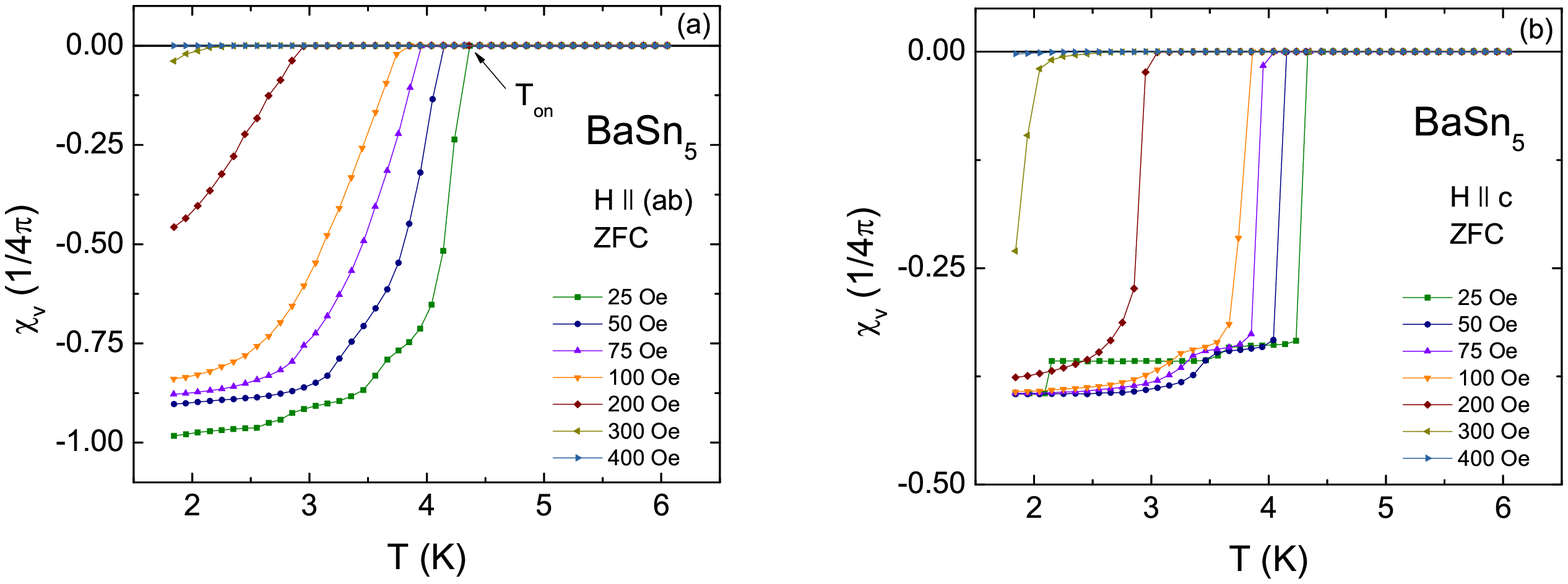}}%
\caption{ZFC temperature-dependent magnetic susceptibility of BaSn$_5$ measured at 25, 50, 75, 100, 200, 300 and 400 Oe. (a) \textit{H}$\parallel$(ab) and (b) \textit{H}$\parallel$c. Criteria for $T_{\rm onset}$ is shwn for the \textit{H}= 25 Oe, \textit{H}$\parallel$(ab) data.}%
\label{Hc2_MT}
\end{center}
\end{figure}

\begin{figure}
\begin{center}
\resizebox*{14cm}{!}{\includegraphics{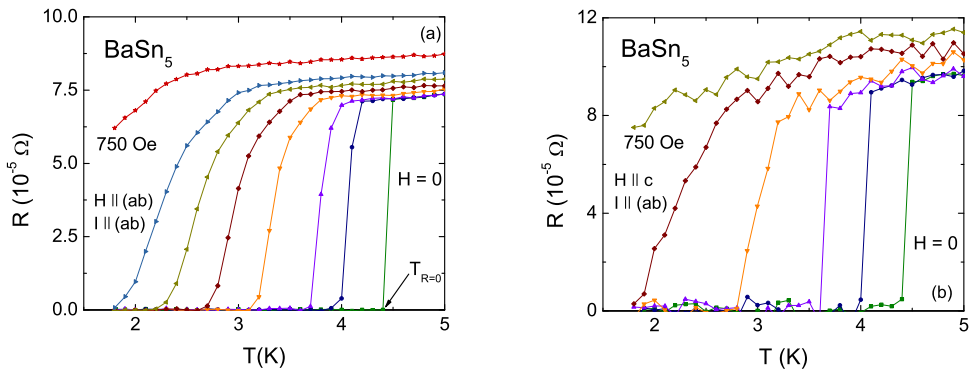}}%
\caption{(a) Low temperature resistance of BaSn$_5$ measured at 0, 50, 100, 250, 500 and 750 Oe with \textit{H}$\parallel$(ab). (b) Low temperature resistance of BaSn$_5$ measured at 0, 50, 100, 200, 300, 400, 500 and 750 Oe with \textit{H}$\parallel$c. Criteria for $T_{R=0}$ is shown for the \textit{H}= 0, \textit{H}$\parallel$(ab) data.}%
\label{Hc2_RT}
\end{center}
\end{figure}

\begin{figure}
\begin{center}
\resizebox*{10cm}{!}{\includegraphics{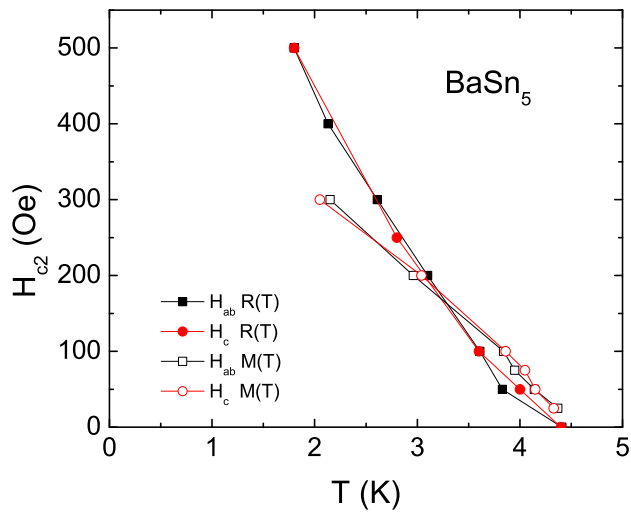}}%
\caption{The upper critical field of BaSn$_5$ from magnetization and magnetotransport measurements.}%
\label{Hc2}
\end{center}
\end{figure}

The zero-field-cooled (ZFC) susceptibilities measured at a set of different low fields are presented in Fig \ref{Hc2_MT} (no corrections for demagnetization factor were employed). At 25 Oe, a sharp superconducting transition is clearly seen with the onset of $\approx$ 4.4 K. To infer an anisotropic upper superconducting critical field for BaSn$_5$, the first data point that deviates from the normal state is chosen as the criterion of superconducting transition. Alternatively, anisotropic $H_{\rm c2}(T)$ can be evaluated from the shifts of resistively measured superconducting transitions in different applied magnetic fields (Fig \ref{Hc2_RT}). \textit{R} = 0 is chosen as the $T_{R=0}$ criteria in resistivity data. Multiple $M(T)$ and $R(T)$ measurements were carried out on different samples, and the data are consistent with each other. The resulting anisotropic $H_{\rm c2}(T)$ curves are shown in Fig \ref{Hc2}, in which $H_{\rm c2}(T)$ obtained from the magnetization data agrees with $H_{\rm c2}(T)$ obtained from the resistivity data quite well. Linear extrapolations yielded a upper critical field of $\sim$ 550 Oe at \textit{T} = 0 K from $M(T)$ measurement, and $H_{\rm c2}(T = 0)$ $\approx$ 950 Oe from $R(T)$ data. Both measurements clearly show that BaSn$_5$ maintains a rather small upper critical field. Despite of the difference in the $H_{\rm c2}$ values, both $M(T)$ and $R(T)$ support that BaSn$_5$ shows almost isotropic behaviour in its superconducting state as seen in Fig \ref{Hc2}. It should be noticed that $T_c$ and superconducting critical fields obtained for BaSn$_5$ in this work are different from that for elemental Sn used as flux ($T_c$(Sn) $\approx$ 3.7 K, and $H_{\rm c2}({\rm Sn}, T= 0)$ = 305 Oe), which rules out traces of Sn flux in the crystals as the source of the superconducting behaviour.

\begin{figure}
\begin{center}
\resizebox*{10cm}{!}{\includegraphics{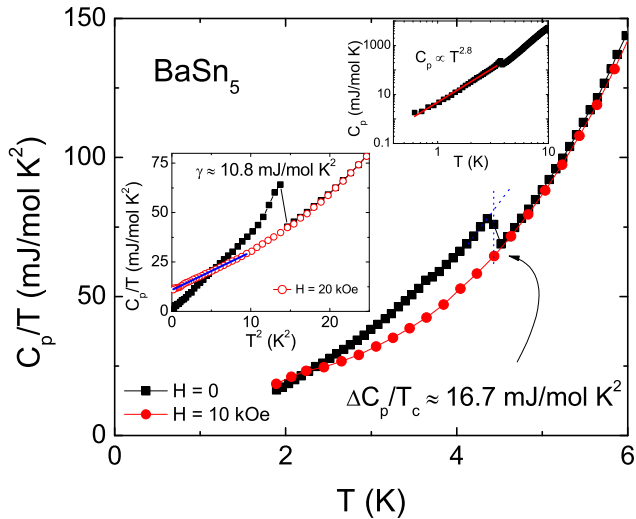}}%
\caption{Low temperature heat capacity of BaSn$_5$ plotted as $C_{\rm p}(T)$ versus \textit{T} in zero and 10 kOe (\textit{H}$\parallel$(ab)) applied field. Lower left inset: Low temperature heat capacity of BaSn$_5$ plotted as $C_{\rm p}(T)$ versus $T^2$ in zero and 20 kOe (\textit{H}$\parallel$(ab)) applied field, solid line -- extrapolation of the low temperature linear region of the 20 kOe data. Upper right inset: zero field $C_{\rm p}(T)$ data plotted on a log-log scale, the solid line corresponds to $C_{\rm p} \propto T^{2.8}$.}%
\label{Cp}
\end{center}
\end{figure}

The low temperature heat capacity of BaSn$_5$ was measured in both zero and applied magnetic field (Fig \ref{Cp}). It is clearly seen that the superconductivity is completely suppressed in 10 kOe without changing its normal state properties. A clear jump at about 4.4 K in the zero-field heat capacity data is associated with the superconducting transition, which gives $\Delta C_{\rm p}/T_{\rm c} \approx$ 16.7 mJ/mol K$^2$. The lower left inset to Fig \ref{Cp} shows the low temperature (down to 0.4 K), in-field (20 kOe), heat capacity data, the Sommerfeld coefficient for BaSn$_5$ is estimated to be $\gamma \approx$ 10.8 mJ/mol K$^2$, and the Debye temperature $\Theta_{\rm D} \approx$ 182.5 K. Thus, $\Delta C_{\rm p}/\gamma T_{\rm c}$ can be estimated to be about 1.55. This value is slightly higher than the canonical 1.43 value expected for isotropic weakly coupled BCS superconductor and suggests that BaSn$_5$ might be strongly coupled superconductor \cite{Carbotte1990}. Finally, the $C_{\rm p}(T)$ behaviour in the superconducting state (Fig \ref{Cp}, upper right inset) appears to be non-exponential and reasonably well described by $C_{\rm p} \propto T^{2.8}$ function. If intrinsic, such dependence might point to deviations from isotropic single band superconductivity for this material.

\begin{figure}
\begin{center}
\resizebox*{10cm}{!}{\includegraphics{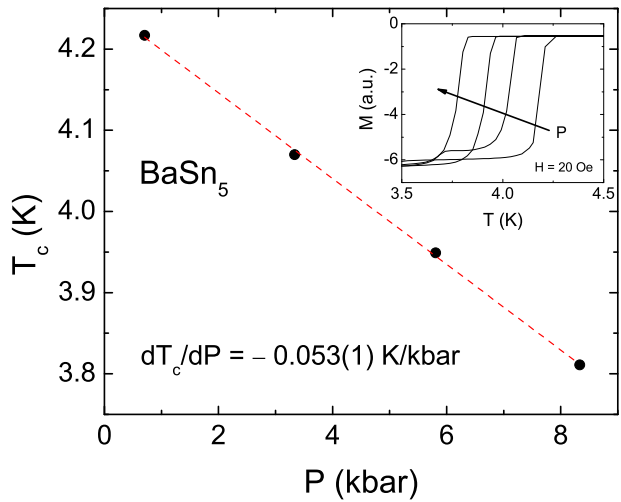}}%
\caption{Pressure dependence of the superconducting transition temperature of BaSn$_5$. Dashed line -- linear fit. Inset: low field magnetization under pressure, the arrow points in the direction of increasing pressure.}%
\label{Pressure}
\end{center}
\end{figure}

The superconducting transition temperature of BaSn$_5$ linearly decreases under pressure up to $\sim$ 8 kbar (Fig \ref{Pressure}). The pressure derivative d$T_{\rm c}$/d$P \approx -0.053$ $\pm$ 0.001 K/kbar, is rather small, similar in sign and order of magnitude to those measured for a number of elemental and binary superconductors \cite{Brant1965}. Such pressure dependence is possibly the result of rather weak dependence of the density of states on energy near the Fermi level as well as possibly opposing changes to $T_{\rm c}$ caused by shift in phonon spectrum by hydrostatic pressure.

\section{Summary}
In this paper we present the synthesis of high quality single crystalline BaSn$_5$, as well as detailed studies on its thermodynamic and transport properties. BaSn$_5$ manifests metallic behaviour in its normal state with (RRR) $\sim$ 1200. Its normal-state magnetic susceptibility is diamagnetic and slightly anisotropic. De Haas-van Alphen oscillations were observed in low temperatures and high fields with the applied magnetic fields parallel to c-axis, two effective masses were resolved via FFT. BaSn$_5$ superconducts at $\sim$ 4.4 K with the upper critical field not exceeding 1 kOe. $H_{\rm c2}$ shows almost isotropic behaviour. $T_{\rm c}$ decreases slowly under hydrostatic pressure up to 10 kbar. The heat capacity data suggest that superconductivity in BaSn$_5$ may be more complex than isotropic BCS. 

Since both Haas-van Alphen oscillation and superconducting state are observed for these high-quality BaSn$_5$ single crystals, detailed study on angular dependence of the oscillatory behaviour, Fermi topology and the symmetry of the superconducting state could be of interest.

\section*{Acknowledgements}
This work was carried out at the Iowa State University and supported by the AFOSR-MURI grant No. FA9550-09-1-0603 (X. Lin and P. C. Canfield). Part of this work was performed at Ames Laboratory, US DOE, under contract No. DE-AC02-07CH 11358 (S. L. Bud'ko).

\end{document}